\documentclass{aastex63}
\usepackage{lineno}
\shorttitle{Variability of the Photospheric Emission}
\shortauthors{Wang et al.}
\usepackage{color}
\newcommand{\MyFigA}{\textcolor[rgb]{0.00,0.00,1.00}{1}}    
\newcommand{\MyFigB}{\textcolor[rgb]{0.00,0.00,1.00}{2}}
\newcommand{\MyFigC}{\textcolor[rgb]{0.00,0.00,1.00}{3}}
\newcommand{\MyFigD}{\textcolor[rgb]{0.00,0.00,1.00}{4}}

\newcommand{\MyTabA}{\textcolor[rgb]{0.00,0.00,1.00}{1}}

\begin{document}
\title{Photospheric Emission in Gamma-ray Bursts: Variability}
\correspondingauthor{Da-Bin Lin}
\email{lindabin@gxu.edu.cn}
\author{Kai Wang}
\affil{Laboratory for Relativistic Astrophysics, Department of Physics, Guangxi University, Nanning 530004, China}
\author{Da-Bin Lin}
\affil{Laboratory for Relativistic Astrophysics, Department of Physics, Guangxi University, Nanning 530004, China}
\author{Yun Wang}
\affil{Laboratory for Relativistic Astrophysics, Department of Physics, Guangxi University, Nanning 530004, China}
\author{Lu-yao Jiang}
\affil{Laboratory for Relativistic Astrophysics, Department of Physics, Guangxi University, Nanning 530004, China}
\author{Shen-Shi Du}
\affil{Laboratory for Relativistic Astrophysics, Department of Physics, Guangxi University, Nanning 530004, China}
\author{Xiao-Yan Li}
\affil{Laboratory for Relativistic Astrophysics, Department of Physics, Guangxi University, Nanning 530004, China}
\author{Jia Ren}
\affil{Laboratory for Relativistic Astrophysics, Department of Physics, Guangxi University, Nanning 530004, China}
\author{Xiang-gao Wang}
\affil{Laboratory for Relativistic Astrophysics, Department of Physics, Guangxi University, Nanning 530004, China}
\author{En-Wei Liang}
\affil{Laboratory for Relativistic Astrophysics, Department of Physics, Guangxi University, Nanning 530004, China}
\begin{abstract}
It is generally believed that the variability of photospheric emission in gamma-ray bursts (GRBs) traces that of the jet power.
This work further investigates the variability of photospheric emission in a variable jet.
By setting a constant $\eta$ (dimensionless entropy of the jet),
we find that the light curve of the photospheric emission shows a ``tracking'' pattern
on the time profile of jet power.
However, the relative variability is significantly low
in the photospheric emission
compared with that in the jet power.
If the $\eta$ is genetic variable,
the variability of the photospheric emission is not only limited
by the jet power but also affected by $\eta$ strongly.
It becomes complex and is generally different from that of the jet power.
Moreover, the opposite phase may stand in the variabilities of the photospheric emission at different photon energies.
We also find that the relative variability does not remain constant over the photon energies with an obvious reduction at a certain energy.
This is consistent with the analysis of GRB 090902B in which an appreciable thermal component has
been detected in a wide energy range. For several other GRBs coupling with the thermal component, we
conservatively evaluate the variability of the thermal and non-thermal emission, respectively.
Our results show that the relative variability of the thermal emission is likely comparable to that of the non-thermal emission for these bursts.
In addition, the analysis of GRB~120323A reveals that the variability of the photospheric emission
may be of the opposite phase from that of the non-thermal emission.
\end{abstract}

\keywords {Gamma-ray bursts (629) --- jets (870) --- Astronomy data analysis (1858) --- Non-thermal radiation sources (1119)}
\section{Introduction}\label{Sec:Intro}
The physical origin of the prompt emission in gamma-ray bursts (GRBs) is still controversial.
Two main categories are divided based on the large statistics of prompt temporal and spectral properties of these energetic events.
One involves the non-thermal emission observed in most of GRBs, and previous works have shown that the synchrotron
or synchrotron-self-Compton (SSC) radiation emitted from accelerated electrons is the promising mechanism (\citealp{1996ApJ...466..768T}; \citealp{2000ApJ...543..722L}; \citealp{2002ApJ...581.1236Z}; \citealp{2011A&A...526A.110D}; \citealp{2011ApJ...726...90Z}; \citealp{2014NatPh..10..351U}).
Another mechanism is a Comptonized quasi-thermal emission from the outflow photosphere
(\citealp{1994MNRAS.270..480T}; \citealp{1999ApJ...511L..93G}; \citealp{2006ApJ...642..995P}; \citealp{2007ApJ...666.1012T}; \citealp{2008A&A...480..305G}; \citealp{2010ApJ...725.1137L};
\citealp{2011ApJ...732...26M}; \citealp{2013ApJ...765..103L}; \citealp{2013ApJ...772...11R}),
according to the quasi-thermal components detected in the spectrum of some GRBs (\citealp{2004ApJ...614..827R}; \citealp{2005ApJ...625L..95R}; \citealp{2009ApJ...702.1211R}; \citealp{2009ApJ...706L.138A};
\citealp{2010ApJ...709L.172R}; \citealp{2011ApJ...730..141Z}; \citealp{2011ApJ...727L..33G}; \citealp{2012ApJ...757L..31A}; \citealp{2013MNRAS.432.3237G}; \citealp{2015ApJ...800L..34L}; \citealp{2013ApJ...770...32G}; \citealp{2011MNRAS.415.1663T}).
Although the thermal components are rarely observed,
their contributions on the GRB prompt emission could not be ignored.
Some authors have suggested that the photospheric emission is an inherent
component for the fireball model (\citealp{2000ApJ...530..292M}; \citealp{2002ApJ...578..812M}; \citealp{2002MNRAS.336.1271D}; \citealp{2005ApJ...628..847R}).
Moreover, several bursts show a distinct thermal component, e.g., GRB 090902B, which highlights
the importance of the photospheric emission during GRB prompt phase.

The particle density at the base of the jet is quite high, which means that the Thomson scattering optical depth is deep enough
so that the photons must be thermal or quasi-thermal.
Thermalized photons are released at the photosphere where the jet becomes transparent to scattering.
\cite{2008ApJ...682..463P} suggested that the position of the photosphere is complicated because the photons can be scattered at any position of the outflow where electrons exist.
The propagating directions and frequencies of photons can be changed in each scattering event, and the observed flux
and temperature of the thermalised photons depend on the last scattering position and angle, scattering time, and co-moving temperature at the last scattered position.
Thus, they proposed a probability density function $P(r, \theta)$ to describe the finite probability of a photon emerging from an arbitrary radius and angle
, i.e., the so-called probability photosphere model.
Different from the traditional position of photosphere where the optical depth equals to unity,
the probability photosphere model pointed out that photons at different positions will escaped to the observer according to the escape probability along the line of sight.
Modifications of this probability density function have been doing based on this model (\citealp{2011ApJ...732...49P}; \citealp{2011ApJ...737...68B}; \citealp{2014ApJ...785..112D}).

The significantly variable light curves of GRB prompt gamma-ray emission are usually accompanied with many overlapping spikes in a short duration or a single smooth large pulse from the observational (\citealp{1995ARA&A..33..415F}).
The variability of the prompt emission has spurred the intense debate of such modes like the collisions of internal shocks (\citealp{1994ApJ...427..708P}; \citealp{1994ApJ...430L..93R};\citealp{1997ApJ...490...92K} ; \citealp{1998MNRAS.296..275D}; \citealp{2009A&A...498..677B}), relativistic mini-jets (\citealp{2003ApJ...597..998L}
\citealp{2004ApJ...607L.103Y}; \citealp{2014ApJ...782...92Z}), and relativistic turbulence (\citealp{2009MNRAS.395..472K}; \citealp{2009ApJ...695L..10L}; \citealp{2009MNRAS.394L.117N}; \citealp{2013ApJ...776...41L}) in a relativistic jet.
However, these models are not applicable to the observed variability in thermal emission which is naturally associated with some inherent properties of the jet, e.g. the jet power and dimensionless entropy.
The observed properties of the thermal emission will vary
when these properties of the jet vary during the burst (\citealp{2013MNRAS.433.2739I}).
The study of thermal emission focuses on energy spectrum and some authors also try to analyze the light curve of thermal emission to find some interesting information.
In this paper, we simulate the variability of the light curve of the photospheric emission and further understand the photospheric emission by the relationship between the variability of the light curve and the intrinsic properties of the jet.
In addition, the exploration of photospheric emission variability may be used as a means of studying thermal emission of GRBs.

The paper is organized as follows.
In Section~\ref{Sec:Model}, we detail the probability photosphere model to calculate the photospheric emission, based on the theoretical work in \cite{2014ApJ...785..112D}.
In Section~\ref{Sec:theory}, we study the variability of photospheric emission in the case of the jet power and dimensionless entropy are both variable.
Following in Section~\ref{Sec:discuss}, some GRB cases in which the thermal components have been confirmed are discussed based on our theoretical results.
The conclusion is placed in Section~\ref{Sec:conclusion}.

\section{Procedure to Calculate the Photospheric Emission}\label{Sec:Model}
In reality, an individual photon in the jet can be scattered to an observer
by an electron at any position in the outflow with a certain probability.
Then, the photospheric emission should be calculated based on a probability density function $P$
(\citealp{2011ApJ...732...49P}; \citealp{2011ApJ...737...68B}; \citealp{2014ApJ...785..112D}; \citealp{2018ApJ...860...72M}; \citealp{2019ApJ...882...26M}).
In spirit of \cite{2014ApJ...785..112D}, the function
\begin{equation}\label{EQ:PROMEGA}
P(r,\theta) = \frac{{{\sigma _{\rm{T}}}n{D^2}e^{ - \tau (r,\theta)}}}{{4\pi A}}
\end{equation}
is adopted to describe the probability of a photon last scattered from $(r, \theta, \phi)$ to an observer.
The detailed information about parameters in Equation~(\ref{EQ:PROMEGA}) are presented as follows:
\begin{itemize}
\item
The $(r, \theta, \phi)$ with $\theta=0$ being along the jet axis is the spherical coordinate
and $r$ is the radius from the central engine of the GRB.
Since we focus on an axisymmetric jet, the effect of $\phi$ is ignored in the present work.
In addition, an on-axis observer is adopted in this paper.

\item
The $\sigma_{\rm T}$ is the Thomson cross-section and $A=2 \eta^2$ with $\eta$ being the dimensionless entropy of the jet is the normalization factor of $P(r,\theta)$.

\item
The $D=1/[\Gamma (1-\beta \cos \theta)]$ is the Doppler factor,
where $\Gamma$ is the bulk Lorentz factor of the jet and
$c\beta=c\sqrt{1-1/\Gamma^2}$ is the jet velocity.
The Lorentz factor of the fireball (jet) is related to $r$ and is read as
\begin{equation}\label{calGamma}
\Gamma = \left\{ {\begin{array}{*{20}{c}}
{r/r_0 ,}&{r < {r_{\rm{s}}},}\\
{\eta,}&{{r \geqslant r_{\rm{s}}},}
\end{array}} \right.
\end{equation}
where $r_0(=10^7~\rm cm)$ is the radius of the central engine and
the saturation radius $r_{\text s}$ is
\begin{equation}\label{Eq:rs}
r_{\text s}=\eta r_0.
\end{equation}

\item
The $\tau$ is the optical depth for photons from $(r, \theta)$ to an observer and is estimated with
\begin{equation}\label{eq:int_tau_B1}
\tau (r,\theta ) = \left\{ {\begin{array}{*{20}{c}}
{\int_r^\infty  {\frac{{[1 - \beta \cos {\theta _l}]{\sigma _{\rm{T}}}n(l)dl}}{{\cos {\theta _l}}}} \;,}&{\theta  \le \pi /2,}\\
{\int_{r\sin \theta }^\infty  {\frac{{[1 - \beta \cos {\theta _l}]{\sigma _{\rm{T}}}n(l)dl}}{{\cos {\theta _l}}}}  + \int_{r\sin \theta }^r {\frac{{[1 - \beta \cos {\theta _l}]{\sigma _{\rm{T}}}n(l)dl}}{{\cos {\theta _l}}}} ,}&{\theta  > \pi /2,}
\end{array}} \right.
\end{equation}
where $\cos{\theta _l} = \sqrt {1 - {{\sin }^2}{\theta _l}}  = \sqrt {1 - {{(r\sin \theta /l)}^2}} $,
$n(r)=\Gamma n'$, $n^{\prime}=\dot M/(4\pi {m_{\rm{p}}}\beta c\Gamma {r^2})$
with $\dot{M}=P_{\rm{jet}}/\eta c^2$
is the electron number density in the comoving frame,
$m_{\text p}$ is proton mass,
and $P_{\rm{jet}}$ is the jet power.

\end{itemize}

With Equation~(\ref{EQ:PROMEGA}), the observed flux at an observer time $t_{\rm obs}$ and photon energy $E$
from a jet shell launched at $[t, t+dt]$
can be estimated with
\begin{eqnarray}\label{eq:f}
{f_{_E} }(t, {t_{{\rm{obs}}}})dt = \frac{1}{{2d_{\rm{L}}^2}}\int_{r_0}^\infty  {  {\int_0^{\theta_{\rm jet}} {\frac{{d\mathop {{N_0}}\limits^. }}{{d\Omega}}}} }E B(E,T)P(r,\theta)\delta (t_{\rm obs} - t- \frac{ru}{\beta c})\sin \theta d\theta dr,
\end{eqnarray}
where $d_{\rm L}=10^{28}\rm{cm}$ is the luminosity distance of the burst,
$\theta_{\rm{jet}}=10 ^\circ$ is the half-opening angle of the jet,
$d\dot{N_0}/d\Omega=P_{\rm jet}(t)dt/(4\pi  \times 2.7k{T_0})$
with $T_0=[P_{\rm{jet}}(t)/(4 \pi r_0^2 a c)]^{1/4}$
is the photon emission rate per unit solid angle, $\delta$ function takes into account emission from different layers ejected at different center engine activity time $t$ with $u=1-\beta \cos \theta$,
and
\begin{equation}\label{PT}
B(E,T) =\frac{E^2}{2.40 (kT)^3}\frac{1}{\exp[-E/({kT})] - 1}
\end{equation}
is the normalized photon distribution of the thermalized photons with a temperature $T=DT'$, i.e.,
$\int_0^{ + \infty } {B(E,T)dE }  = 1$.
The comoving temperature $T'$ is given by
\begin{equation}\label{T'}
T' = \left\{ {\begin{array}{*{20}{c}}
{{T_0}/(2\Gamma ),}&{r < {r_{\rm{s}}} < {r_{{\rm{ph}}}},}\\
{T_0{{(r/{r_{\rm{s}}})}^{ - 2/3}}/(2\Gamma ),}&{{r_{\rm{s}}} < r < {r_{{\rm{ph}}}},}\\
{{T_0}{{({r_{{\rm{ph}}}}/{r_{\rm{s}}})}^{ - 2/3}}/(2\Gamma ),}&{{r_{\rm{s}}} < {r_{{\rm{ph}}}} < r,}
\end{array}} \right.
\end{equation}
and the photospheric radius $r_{\text{ph}}$ is given by
\begin{equation}\label{Eq:Rph}
r_{\text{ph}}=\frac {1}{(1+\beta) \beta \eta^2} \frac{\sigma_{\text T}}{m_{\text p} c} \frac{P_{\text{jet}}}{4 \pi c^2 \eta}.
\end{equation}
One should note that the photospheric emission is not necessarily  expected to be quasi-thermal.
This is owing to that any dissipation of energy close to the location of the photosphere
will distort and broaden the spectrum of the photospheric emission
(\citealp{2005ApJ...635..476P}, \citealp{2006ApJ...642..995P}; \citealp{2006A&A...457..763G}; \citealp{2006A&A...450..887G}; \citealp{2011ApJ...737...68B}).
For simplify, the dissipation beneath the photosphere is not involved in our work.
This may be applicable for our discussing bursts (see Section~\ref{Sec:discuss}), e.g., GRB~090902B, which can be well fitted with a thermal component.

With Equation~(\ref{eq:f}), the observed total flux $F_E (t_{\rm obs})$ from a continuous jet
can be estimated with
\begin{eqnarray}\label{eq:Fnu}
F_{_E}  (t_{\rm obs}) = \int_0^{{t_{{\rm{obs}}}}} {{f_{_E}  }(t,{t_{{\rm{obs}}}})dt},
\end{eqnarray}
where $t_{\rm obs}=0$ is set at the observed time of photons from the jet flow launched at $t=0$ and location of $r_0$.
The corresponding luminosity $L_{\rm ph}(t_{\rm obs})$ can be given
by integrating Equation~(\ref{eq:Fnu}), i.e.,
\begin{equation}\label{Lt}
L_{\rm ph}({t_{{\rm{obs}}}}) = 4\pi d_{\rm{L}}^2\int_0^{ + \infty } {{F_{_E} (t_{\rm obs})}dE } .
\end{equation}

\section{Variability of photospheric emission}\label{Sec:theory}
\subsection{Theoretical Expectation}\label{Sec:predict}
For a jet with $r_{\text{ph}}>r_{\text s}$, the photospheric luminosity can be estimated with
\begin{equation}\label{Eq:P}
L_{\rm ph} \propto P_{\text{jet}} (r_{\text{ph}}/r_{\text s})^ {-2/3}.
\end{equation}
According to Equations~(\ref{Eq:rs}) and (\ref{Eq:Rph}), Equation~(\ref{Eq:P}) can be reduced to
\begin{equation}\label{Eq:Lph}
L_{\text{ph}} \propto P_{\text{jet}}^{1/3}\eta^{8/3}.
\end{equation}
Then, one can derive the following relation:
\begin{equation}\label{Eq:rms}
{\sigma _{\rm{ph}}} \simeq \sqrt {{{\left( {\sigma _{\rm jet}/3} \right)}^2} + {{\left( {8{\sigma _\eta }/3} \right)}^2}}
\end{equation}
where
the relative root-mean-square amplitude ($\sigma$) of the light curve
is used to describe the variability,
and $\sigma _{\rm{ph}}$, $\sigma _{\rm jet}$, and $\sigma _\eta$ are the values corresponding to
$L_{\rm ph}$, $P_{\text{jet}}$, and $\eta$, respectively.
From Equations~(\ref{Eq:Lph}) and (\ref{Eq:rms}),
one can see that the relative amplitude of the photospheric emission variability
would be $\sim 1/3$ of that of the jet power if $\eta$ is free of variability.
If the $\eta$ is genetic variable, the variability of the photospheric emission is not only limited
by the jet power but also strongly affected by $\eta$.

We also consider the variability of photospheric emission in the low-energy regime.
Theoretically, the emission of photosphere at photon energy $E$ can be read as
\begin{equation}\label{Eq:Energy}
{F_{_E} }dE={L_{{\rm{ph}}}}H(\frac{{E}}{{{E_{\rm{p}}}}})d\left( {\frac{E}{{{E_{\rm{p}}}}}} \right)
\end{equation}
with $\int_0^\infty  {H(x)dx}  = 1$ and $E_{\rm p}$ being the peak photon energy of the $\nu F_{\nu}$ radiation spectrum.
The shape of $H(x)$ can be found in the figure~9 of \cite{2014ApJ...785..112D}.
According to the figure~9 of \cite{2014ApJ...785..112D}, the photospheric emission at low-energy regime, e.g., $E\sim 10^{-3}E_{\rm p}$, can be estimated with
\begin{equation}
{F_{_E}  }d\nu \simeq {L_{{\rm{ph}}}}\frac{{E}}{{{E_{\rm{p}}}}}d\left( {\frac{{E}}{{{E_{\rm{p}}}}}} \right)
\end{equation}
or
\begin{equation}\label{Eq:F_nu_E_p}
{F_{_E}  } \propto {L_{{\rm{ph}}}}E_{\rm p}^{-2}.
\end{equation}
We note that the $E_{\rm p}$ is related to the observed temperature $T$ at the photospheric radius with $E_{\rm{p}} \sim kT$.
Then, Equation~(\ref{Eq:F_nu_E_p}) can be reduced to
\begin{equation}\label{Eq:LowE}
{F_{_E} } \propto P_{\rm jet}^{7/6}{\eta ^{ - 8/3}}.
\end{equation}
It reveals that the relative variability of the photospheric emission
at photon energy $E\sim 10^{-3}E_{\rm p}$ is ${\sigma_{_E} } \simeq \sqrt {{{\left( {7\sigma _{\rm jet}/6} \right)}^2} + {{\left( {8{\sigma _\eta }/3} \right)}^2}}$,
which is the same as $\sigma_{\rm ph}$ if $P_{\rm jet}$ is free of variability.
Moreover, Equation~(\ref{Eq:LowE}) reveals that
the variability of the light curve in low-energy bands may be of the opposite phase from the $\eta$ and $L_{\rm ph}$.
That is to say, the opposite phase may stand in the variabilities of the photospheric emission at different photon energies.

For a jet with $r_{\text{ph}}<r_{\text s}$,
one can have ${L_{{\rm{ph}}}} \propto {P_{{\rm{jet}}}}$ and
$\sigma_{\rm{ph}}=\sigma_{\rm{jet}}$.
That is to say, the variabilities of the photospheric emission are only related
to the variabilities of $P_{\rm{jet}}$ without associations with the variabilities of the $\eta$.
In this work, we explore the variability of photospheric emission from a jet with $r_{\text{ph}}>r_{\text s}$.

\subsection{Calculation Results}\label{Sec:Calculations}
In this section, we numerically calculate the variabilities of photospheric emission.
We first study the variabilities of the photospheric emission with a variable $P_{\rm{jet}}$ only.
Here, the variability of jet power in GRBs is modelled
based on the model of propagating fluctuations in the accretion flow,
which has been used to explain the variabilities of GRBs (\citealp{2016MNRAS.463..245L}).
The light curves of jet power are shown in Figure~{\MyFigA} with black lines,
where the maximum of jet power is set as $10^{52}\,\rm erg\cdot s^{-1}$
and $\sigma_{\rm jet}=0.1$ ($0.3$) is adopted in the left (right) panel.
Then, we calculate the photospheric emission with the aforementioned model in Section~\ref{Sec:Model}
and $\eta=250$.
The obtained light curves of photospheric emission are shown in Figure~{\MyFigA} with red lines.
In Figure~{\MyFigA}, all curves are normalized with their maximal values.
It can be found that the variability of the photospheric emission shows
a ``tracking" pattern on the variability of the jet power.
However, the relative variability of the photospheric emission
is significantly low compared with that of the jet power.
These behaviors are consistent with Equations~(\ref{Eq:Lph}) and (\ref{Eq:rms}).
For better insight,
we also plot the theoretical photospheric luminosity, i.e., Equations~(\ref{Eq:Lph}) with $\eta=250$,
in Figure~{\MyFigA} with cyan lines.
One can find that the variability of photospheric luminosity is almost the same as that of $P_{\rm jet}^{1/3}$,
which is consistent with Equation~(\ref{Eq:Lph}).

Then, we study the effect of $\eta$ variability on the photospheric emission.
The variability of $\eta$ is also modelled based on the model of propagating fluctuations
but it is independent of the jet power variability.
In Figure~{\MyFigB},
we show the light curves of $P_{\rm jet}$ and $\eta$ with gray and cyan lines, respectively.
Here, a quasi-steady jet rather than a fast rise and exponential decay jet is adopted.
Different variabilities of $P_{\rm jet}$ and $\eta$ are adopted in each panel,
i.e., ($\sigma_{\rm jet}, \sigma_{\rm \eta}$)=($0, 0.1$), ($0, 0.2$), ($0.15, 0.1$), ($0.15, 0.2$), ($0.3, 0.1$),
and ($0.3, 0.2$) for the panel (a), (b), (c), (d), (e), and (f), respectively.
Then, we calculate the photospheric emission based on the procedure presented in Section~{\ref{Sec:Model}}.
The obtained light curve of photospheric emission is shown with red line in each panel.
In Figure~{\MyFigB}, all curves are normalized with their arithmetic mean values.
Obviously, the light curve of photospheric emission is very different from that of jet power $P_{\rm jet}$
in a jet with variable $\eta$.
 We estimate the Pearson correlation coefficient $k$
\footnote{
A Pearson correlation coefficient is a number between -1 and 1 that indicates the extent to which two variables are linearly related.
For two sets of data $A (= \left\{ {{A_1},{A_2}, \cdots ,{A_{N - 1}},{A_N}} \right\})$ and $B(= \left\{ {{B_1},{B_2}, \cdots ,{B_{N - 1}},{B_N}} \right\})$,
the Pearson correlation coefficient $k$ is estimated with
${k} = {{{1} \over {N-1}} \sum_{i=1}^{N} ({ {A_{i}- \mu_{A}} \over {\chi_{A}}})({ {B_{i}- \mu_{B}} \over {\chi_{B}}})}$,
where $\mu$ and $\chi$ are the mean value and standard deviation of the data set, respectively.
The situation with $k=0$ indicates that there is no association between the two variables,
$k>0$  indicates a positive association,
and $k<0$ indicates a negative association.
If $| k |$ is close to 1, there would be a strong correlation between the two variables.
}
between $P_{\rm{jet}}$ and $L_{\rm{ph}}$.
For the situations with ($\sigma_{\rm jet}, \sigma_{\rm \eta}$)=($0.15, 0.1$) and ($0.15, 0.2$),
the values of $k=0.1480$ and $0.0460$ are obtained, respectively;
for the situations with ($\sigma_{\rm jet}, \sigma_{\rm \eta}$)=($0.3, 0.1$) and ($0.3, 0.2$),
the values of $k=0.3194$ and $0.1309$ are obtained, respectively.
These results reveal that the correlation between $L_{\rm{ph}}$ and $P_{\rm{jet}}$ increases with the variability of jet power and decreases with the variability of dimensionless entropy.
This behavior can be understood with Equation~(\ref{Eq:Lph}),
which describes the dependence of $L_{\rm ph}$ on both $P_{\rm jet}$ and $\eta$.
Then, we plot $P_{\text{jet}}^{1/3}\eta^{8/3}$ in each panel with green dashed lines.
One can see that the results of the probability photosphere model (red lines)
are well coincident with Equation~(\ref{Eq:Lph}).
Here, the Pearson correlation coefficient between $L_{\rm{ph}}$ and $P_{\text{jet}}^{1/3}\eta^{8/3}$ in the subplots of Figure~{\MyFigB} are all around 0.999.
That is to say, Equation~(\ref{Eq:Lph}) well describes the photospheric emission even for a variable jet.
We calculate the relative amplitude of the photospheric emission variability
and the results (i.e.,  $\sigma_{\rm{ph,obtained}}$) are shown in the bottom of each panel.
This value is close to the result of ${\sigma _{{\rm{ph}}}} \simeq \sqrt {{{\left( {{\sigma _{{\rm{jet}}}}/3} \right)}^2} + {{\left( {8{\sigma _\eta }/3} \right)}^2}}$.
We also study the variability of photospheric emission by simulating the variability of $\eta$
with a serial of random numbers.
It is found that the variability of $L_{\rm ph}$ well traces the variability of $P_{\text{jet}}^{1/3}\eta^{8/3}$.

In the upper panel of Figure~{\MyFigB}, we shows the light curves ($F_E$) of
flux density observed at photon energy $E=10^{-3}E_{\rm p}$ with blue lines.
Obviously, the blue lines have opposite phases with $L_{\rm{ph}}$,
where $k=-0.8168$ and $-0.6412$ are obtained for the $L_{\rm{ph}}$ and $F_E$
in the left and right panels, respectively.
This behavior is consistent with Equation~(\ref{Eq:LowE}).
In Figure~{\MyFigC},
we plot the dependence of $\sigma_{_E}$ on the observed photon energy $E$ for the emission from photosphere. Here, $\sigma_{_E}$ is the relative root-mean-square of the variability in the light curves
observed at photon energy $E$.
The results show that there is a distinct groove in the $\sigma_{_E}-E$ relation
but the position and depth of grooves may be different in jets with different ($\sigma_{\rm{jet}}, \sigma_{\rm{\eta}}$).

\section{Case Discussion}\label{Sec:discuss}
The variabilities of the photospheric emission provide a new approach to explore the properties of thermal emission
detected in GRB prompt spectrum.
In this section, we discuss the variabilities of GRBs' light curves based on several GRBs in which the thermal component has been detected.
We focus on two situations:
\begin{itemize}
\item
A burst dominated by a thermal component.
GRB~090902B is a bright, long gamma-ray burst, detected by the Gamma-ray Burst Monitor (GBM) and Large Area Telescope (LAT) on-board the \emph{Fermi Gamma-ray Space Telescope}.
Some works show that the gamma-ray spectrum of this burst is dominated by the thermal emission in the energy
range from $\sim 50\,\rm{keV}$ to $\sim 10\,\rm{MeV}$ (\citealp{2009ApJ...706L.138A}; \citealp{2010ApJ...709L.172R}; \citealp{2011ApJ...730..141Z}).
We study the relation between the variability amplitudes of light curves and the photon energy in this burst.
In order to reduce the influence of power-law radiation component, we extract the light curves in the energy band $91-1630~\rm{keV}$ and evenly divide into 12 energy bands in the logarithmic space.
The variability amplitudes of the light curves are calculated in these 12 energy bands, which is shown in Figure~{\MyFigC} with the symbol ``$\bigstar$''.
The pattern of $\sigma_{_E}$-$E$ relation resembles our theoretical results.
Note that each energy band is represented by its mean value.
Moreover we fit these light curves polynomially as their mean flux density
when the variability amplitude $\sigma$ is calculated.
\item
The bursts coupling with thermal components, e.g., GRBs~100724B,~110721A,~110920A, and~120323A.
Their spectra is dominated by the typical Band function, which is usually taken to represent a non-thermal emission component, but also
includes a significant thermal spectral contribution
(\citealp{2011ApJ...727L..33G}; \citealp{2012ApJ...757L..31A}; \citealp{2012grb..confE..12M}; \citealp{2013ApJ...770...32G}).
According to the spectral analysis of these papers, we select three energy bands for our discussion.
For GRB~100724B, (I) $8-30~\rm{keV}$, (II) $80-120~\rm{keV}$ and (III) $300-8000~\rm{keV}$;
for GRB~110721A, (I) $10-20~\rm{keV}$, (II) $60-200~\rm{keV}$ and (III) $500-2000~\rm{keV}$;
for GRB~110920A, (I) $8-20~\rm{keV}$, (II) $50-300~\rm{keV}$ and (III) $800-2000~\rm{keV}$;
for GRB~120323A, (I) $8-15~\rm{keV}$, (II) $20-150~\rm{keV}$ and (III) $200-3000~\rm{keV}$ are adopted.
Here, I band and III band are almost contributed from the non-thermal component,
and II band is dominated by the thermal component.
Firstly, we obtained the light curves of each energy band.
The result is shown in the left panel of Figure~{\MyFigD}, where only GRB~120323A is demonstrated as an example.
One can intuitively see that the variability of II band is weak by comparing with that of the non-thermal emission.
Then, we estimate the relative variability amplitude ($\sigma$) of the light curves in the above energy bands.
The values of $\sigma$ and $\Re=\sigma_{_{\rm{II}}}/\sigma_{_{\rm{III}}}$ are reported in Table~{\MyTabA}.
It can be found that the relative variability amplitudes of the light curves in II band are slightly lower than that in I or III band for these bursts.

$\;\;\;\;\;\;$
In fact, the spectrum in each energy band is a superposition of the thermal and non-thermal segments, thus our estimations are conservative.
The spectra in bands I and III can be treated to be completely contributed by the non-thermal component.
But neither of the components can be ignored in the energy band II.
By fitting the time-integrated spectra of these bursts,
we estimate the total flux ($F_{\rm{II}}$) and the flux from the photosphere ($F_{\rm{ph}}$) in the energy band II.
The fraction of the photospheric emission in band II can be evaluated with $f_{\rm{ph}}=F_{\rm{ph}}/F_{\rm{II}}$,
which is reported in Table~{\MyTabA}.
The parameter $f_{\rm{ph}}$ can help us to understand the variability of photospheric emission
as follows.
If the variabilities of the photospheric and non-thermal emission are independent,
the variability of light curve in energy band II would be subject to
$\sigma_{\rm{II}}=\sqrt{\sigma_{\rm{ph}}^2 f_{\rm{ph}}^2+\sigma_{\rm{nth}}^2 (1-f_{\rm{ph}}^2)}$.
Here, $\sigma_{\rm ph}$, $\sigma_{\rm nth}(=\sigma_{\rm III})$, and $\sigma_{\rm II}$
are the relative variability of the photospheric emission, the non-thermal emission,
and the hybrid one, respectively.
With this formula, we simulate the relation of $\Re$ on $f_{\rm{ph}}$ which is shown in the right panel of Figure~{\MyFigD} and our bursts are also shown with symbols ``$\Delta$''.
From these results, we can conclude that,
(1) The ratio of GRB~120323A variability between thermal and non-thermal component is significantly lower than the minimum value ($\sigma_{\rm ph}=0$) in our simulation.
Then, the variability of the photospheric emission should be associated with that of the non-thermal emission in GRB~120323A.
In addition, the phase of variability in the photospheric emission should be inverse with respect to that in the non-thermal emission component.
This behavior can be found in the light curves shown in the left panel of Figure~{\MyFigD}.
(2) The variability of the photospheric emission in GRBs~100724B,~110721A, and~110920A
should be comparable to that of the non-thermal component, i.e., $\sigma_{\rm ph}\sim \sigma_{\rm III}$.
In the photosphere-internal shock scenario,
the non-thermal emission may be from the internal shocks,
which is formed in a jet with highly variable dimensionless entropy.
If the variability of photospheric emission is dominated by that induced by the variable dimensionless entropy,
our results reveal that the relative variability ($\sigma_{_{\rm IS}}$) of observed emission from internal shock is almost 3-4 times of that of dimensionless entropy, i.e., $\sigma_{_{\rm IS}}\sim 3\sigma_{\eta}- 4\sigma_{\eta}$.
\end{itemize}

\section{Conclusion and Discussion}\label{Sec:conclusion}
In this paper, we study the variability of the photospheric emission in GRBs, with both the effects
of the jet power $P_{\rm{jet}}$ and dimensionless entropy $\eta$ have been considered.
By considering a constant dimensionless entropy, a good ``tracking'' pattern,
i.e., the light curves of photospheric emission tracking the time profile of the jet power,
is obtained in our calculations.
However, the relative variability of photospheric emission is significantly
low compared with that of the photospheric emission.
We then pay attention on the variability of photospheric emission
from a jet with variable $P_{\rm{jet}}$ and variable $\eta$.
It is found that the variability of $\eta$ has a stronger effect on the variability of the photospheric emission than that of $P_{\rm{jet}}$.
With a variable $\eta$, the variability of the photospheric emission becomes complex
and can be very different from that of the jet power.
Moreover, it shows opposite phase for the variability of the photospheric luminosity
and that of the photospheric emission at low photon energies
if the effect of a variable jet power on the variability of photospheric emission can be ignored
compared with that of a viable $\eta$.
We also study the dependence of the relative variability on the observed photon energy.
The amplitude of relative variability remains constant over the photon energy
and has an obvious reduction near a certain photon energy.
This result is coincident with the analysis of GRB~090902B,
in which an appreciable thermal component has been detected in a wide energy range.
Analyses of GRBs~100724B, 110721A, and 110920A show
that the amplitude of the relative variability in the photospheric emission
is likely comparable to that in the non-thermal emission for these bursts.
Moreover, the analysis of GRB~120323A shows that
the variability of the photospheric emission is related to that of the non-thermal emission
with phase deviation in their variabilities.

In fact, in addition to the matter-dominated fireball model, the jet may be highly magnetized.
In a magnetically dominated jet, the energy dissipation caused by magnetic reconnection will affect the properties of photospheric emission (\citealp{2002A&A...387..714D}, \citealp{2002A&A...391.1141D}).
\cite{2007A&A...469....1G} provided a formula to describe the luminosity of the photosphere
in a magnetically dominated jet, i.e.,
\begin{equation}\label{Eq:L_ph_magnetized}
L_{\rm{ph}} \propto {P_{{\rm{jet}}}^{6/5}} \zeta_0^{-3/2},
\end{equation}
where $\zeta_0$ is the initially magnetization of the jet.
Equation~(\ref{Eq:L_ph_magnetized}) is obtained by setting $r_{\rm{ph}} < r_{\rm{s}}$.
For a highly magnetized jet,
${r_{\rm{s}}} = \pi c\Gamma _\infty^2(3\varepsilon \Omega ) ^{-1}$
and
${r_{{\rm{ph}}}} = 6 \times {10^{ - 16}}P_{{\rm{jet}}}^{3/5}{( \varepsilon \Omega )^{-2/5}}\zeta _0^{-3/2}$,
where $\Gamma_{\rm \infty}$ is the terminal bulk Lorentz factors for the jet,
$\Omega$ is the angular frequency of the compact object in the GRB central engine,
$\varepsilon$ is the ratio of the magnetic reconnection velocity to the Alfv\'{e}n speed (\citealp{2007A&A...469....1G}).
Then, the variability of the photospheric emission can be estimated with ${\sigma _{\rm{ph}}} \simeq \sqrt {{{\left( {6 \sigma _{\rm jet}}/5 \right)}^2} + {{\left( {3{\sigma _{\zeta_0}}/2} \right)}^2}}$.
Note that the terminal bulk Lorentz factors $\Gamma_{\rm \infty}$ of a matter-dominated and magnetically dominated jets
are related to $\eta$ and $\zeta_{0}$, i.e., $\Gamma_{\rm \infty}= \eta$ and $\Gamma_{\rm \infty}= \zeta_{0}^{3/2}$, respectively.
For comparison, we would like to replace $\eta$ and $\zeta_{0}$ with $\Gamma_{\rm \infty}$.
Equations~(\ref{Eq:Lph}) and (\ref{Eq:L_ph_magnetized}) can be reduced to
$L_{\rm{ph}} \propto {P_{{\rm{jet}}}^{1/3}} \Gamma_{\rm \infty}^{8/3}$ and
$L_{\rm{ph}} \propto {P_{{\rm{jet}}}^{6/5}} \Gamma_{\rm \infty}^{-1}$, respectively.
Then, the variabilities of the photospheric emission can be estimated with
${\sigma _{{\rm{ph}}}} \simeq \sqrt {{{\left( {{\sigma _{{\rm{jet}}}}/3} \right)}^2} + {{\left( {8{\sigma _{{\Gamma _{{\rm{\infty}}}}}}/3} \right)}^2}}$
and
${\sigma _{{\rm{ph}}}} \simeq \sqrt {{{\left( {6{\sigma _{{\rm{jet}}}}/5} \right)}^2} + {{\left( {{\sigma _{{\Gamma _{{\rm{\infty}}}}}}} \right)}^2}}$, respectively.
It reveals that the variability of photospheric emission
in a magnetically dominated jet has a stronger (weaker) dependence on the variability of $P_{\rm jet}$ ($\Gamma_{\rm \infty}$) compared with that in a matter-dominated jet.
In addition, $L_{\rm{ph}}$ is anticorrelated with $\Gamma_{\rm \infty}$ in a magnetically dominated jet,
which is very different from that in a matter-dominated jet.
In a matter-dominated jet, the non-thermal emission may be produced in the internal shocks,
which is related to the fluctuations of Lorentz factors.
In a magnetically dominated jet,
the relation between the thermal emission and non-thermal emission may be complex.
If $r_{\rm{ph}}\ll r_{\rm s}$,
the amount of energy dissipated above the photosphere,
which can potentially power the non-thermal component,
scales roughly like $P_{\rm jet}$
and is unrelated to $\Gamma_{\rm \infty}$.
In this situation,
the variability of the non-thermal emission would almost proportional to
the variability of the thermal emission if ${\sigma _{{\Gamma _\infty }}}/{\sigma _{{\rm{jet}}}}\ll1$.
However, the variability of the non-thermal emission may be independent on the variability of the thermal emission if ${\sigma _{{\Gamma _\infty }}}/{\sigma _{{\rm{jet}}}}\gg1$.
This behavior is very different from that in a matter-dominated jet.

\acknowledgments
We thank the anonymous referee for beneficial suggestions that improved the paper.
We thank Xue-Feng Wu and Yan-Zhi Meng for helpful discussions and suggestions.
This work is supported by the National Natural Science Foundation of China (grant Nos.11773007, 11533003, 11851304, U1731239), the Guangxi Science Foundation
(grant Nos. 2018GXNSFFA281010, 2016GXNSFDA380027, 2017AD22006, 2018GXNSFGA281005), and the Innovation Team and Outstanding Scholar Program in Guangxi Colleges.

\clearpage
\begin{figure}
\centering
\begin{tabular}{cc}
\includegraphics[angle=0,width=0.5\textwidth]{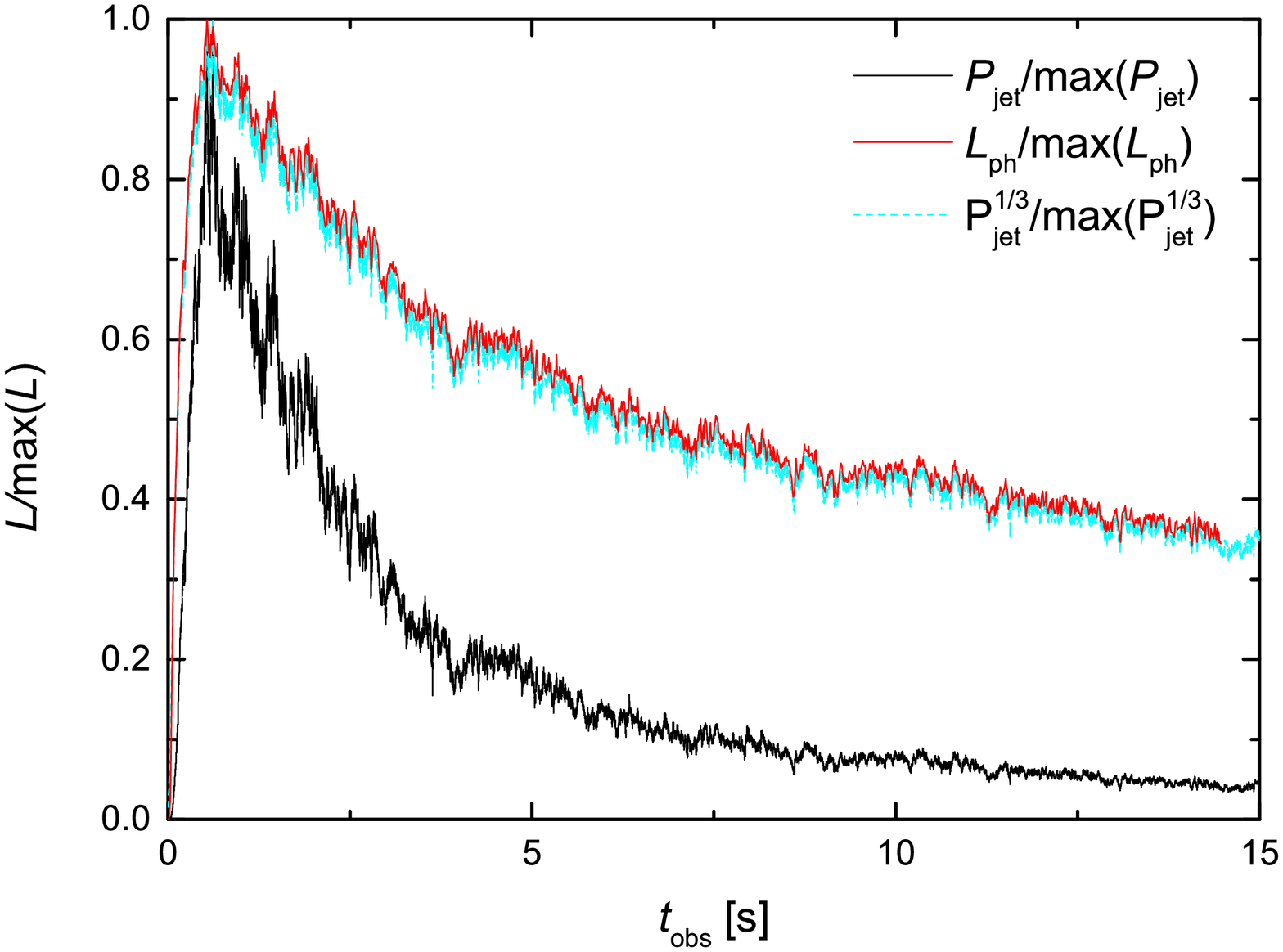} &
\includegraphics[angle=0,width=0.5\textwidth]{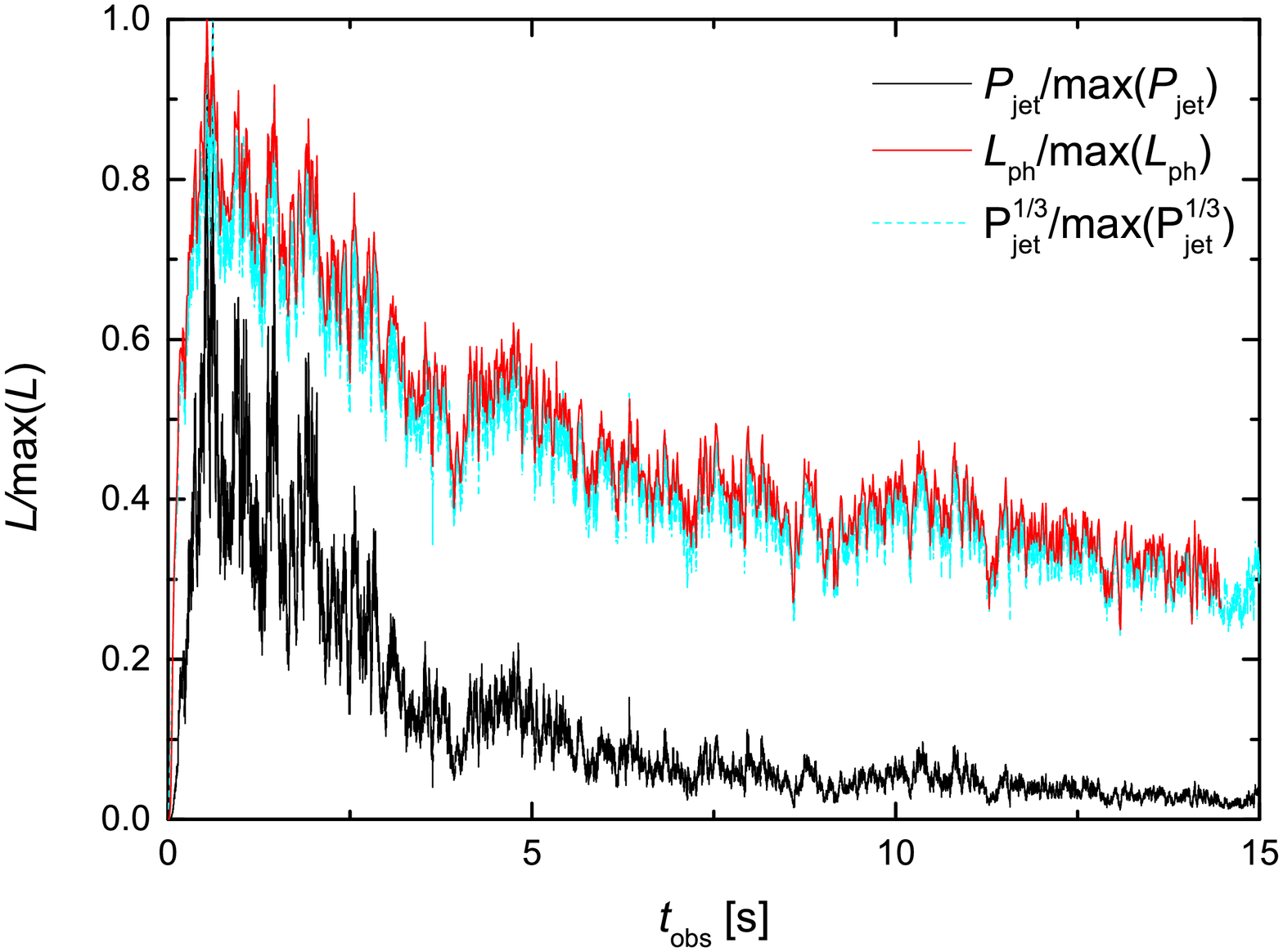} \\
\end{tabular}
\caption{Light curves of $P_{\rm{jet}}$ (black lines), $L_{\rm ph}$ (red lines), and $P_{\rm{jet}}^{1/3}$  (cyan lines),
where jets with $\sigma_{\rm{jet}}=0.1$ (left panel) and $\sigma_{\rm{jet}}=0.3$ (right panel) are studied.}
\label{fig:L1}
\end{figure}

\clearpage
\begin{figure}
\centering
\begin{tabular}{cc}
\includegraphics[angle=0,width=0.5\textwidth]{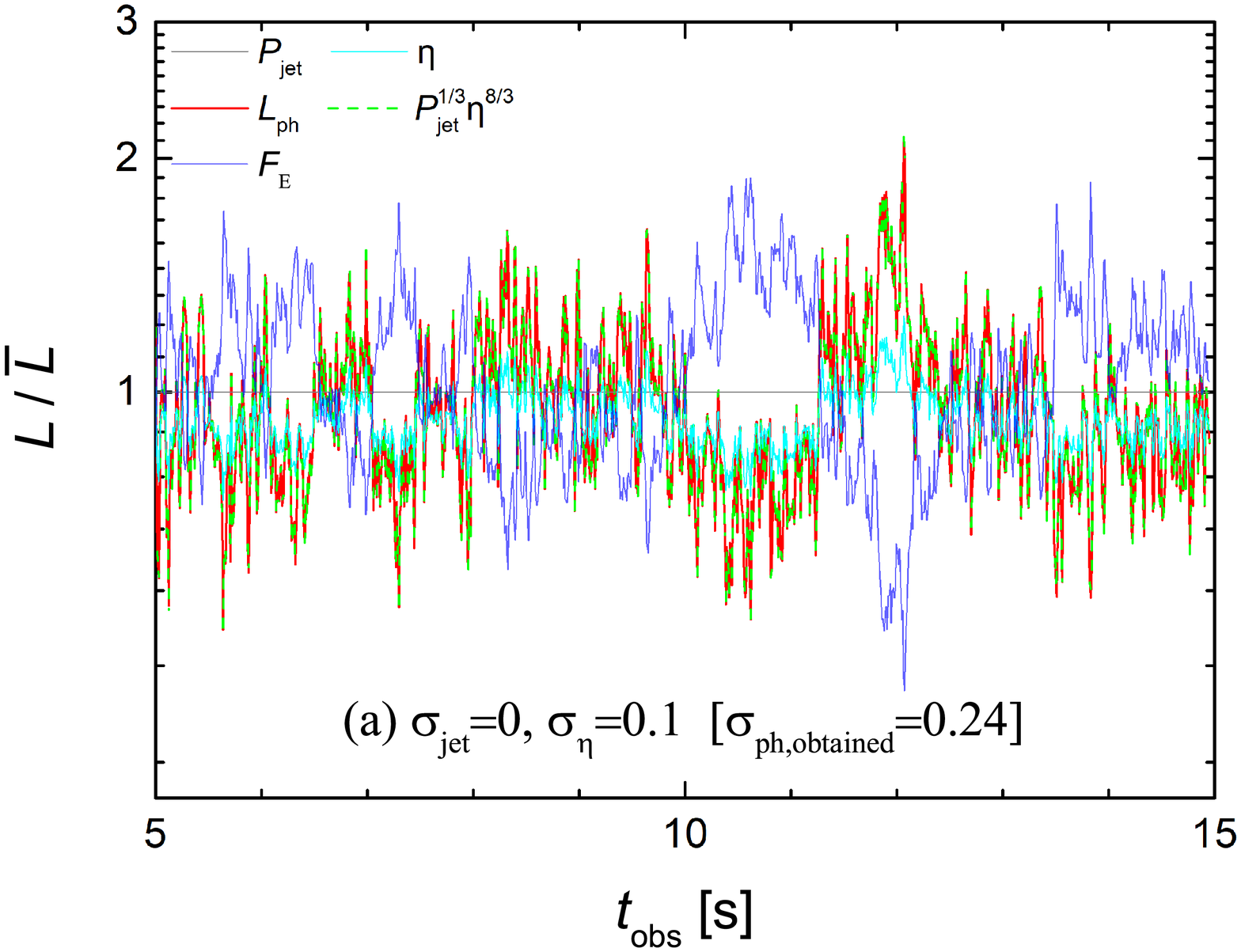} &
\includegraphics[angle=0,width=0.5\textwidth]{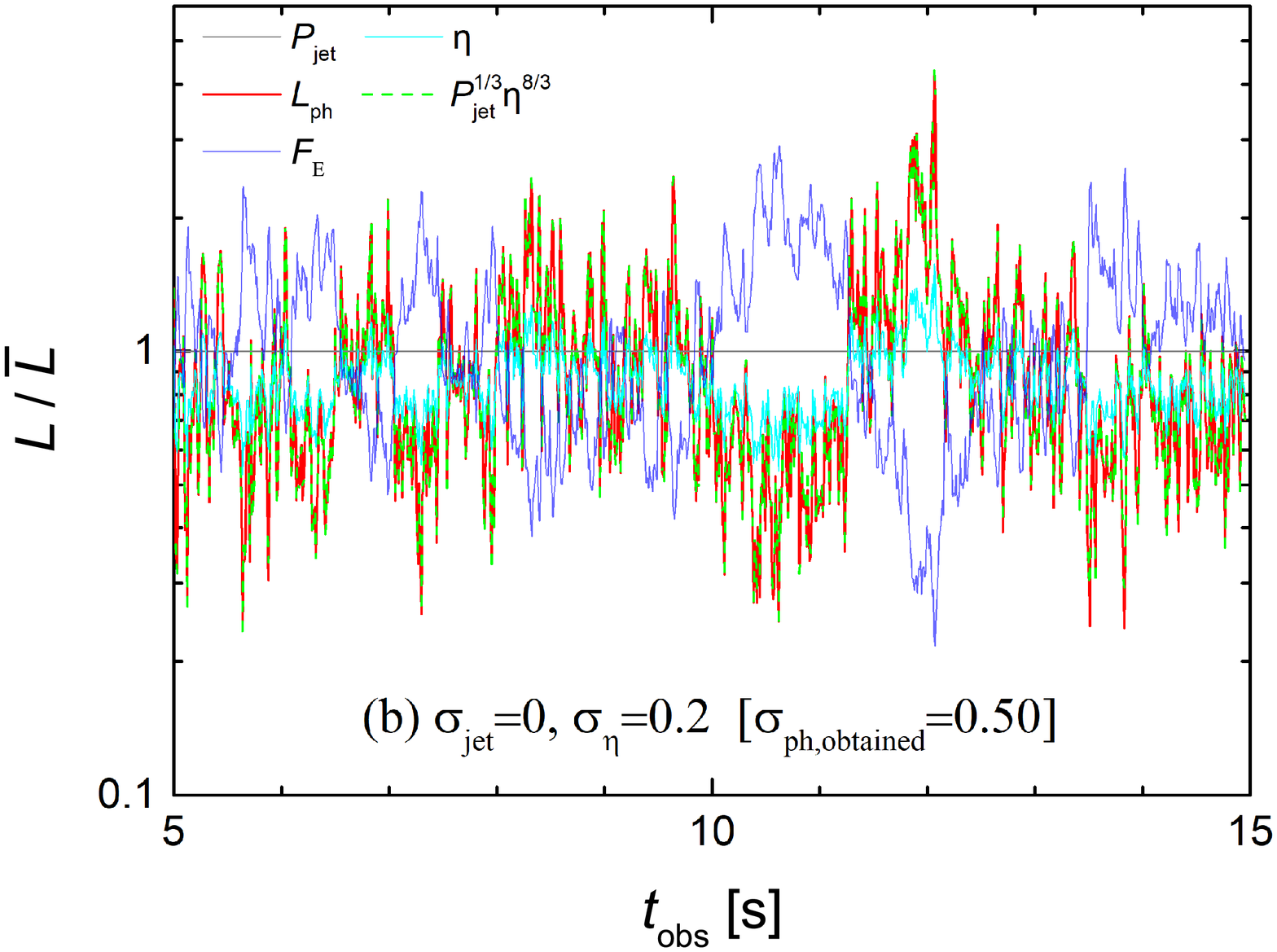} \\
\includegraphics[angle=0,width=0.5\textwidth]{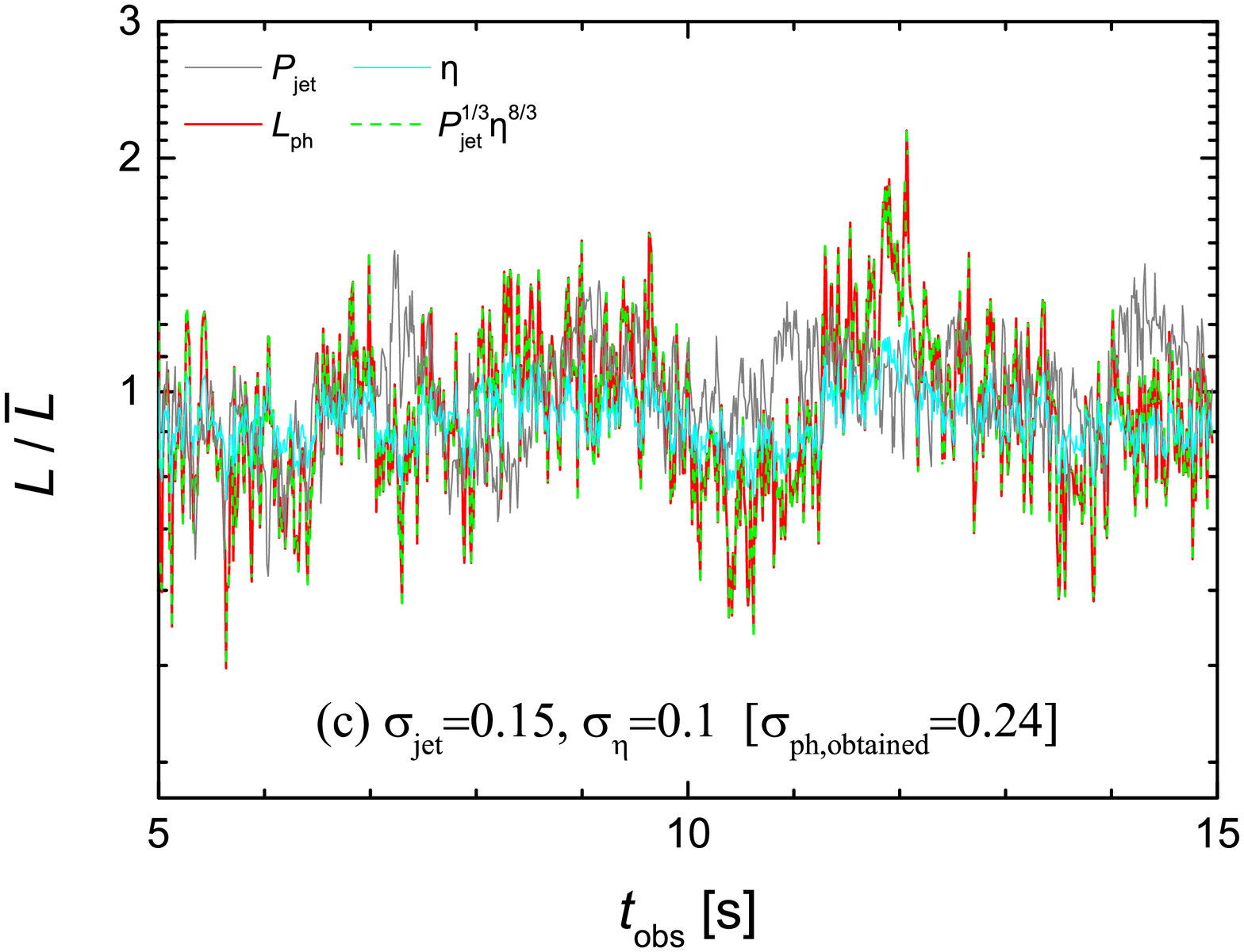} &
\includegraphics[angle=0,width=0.5\textwidth]{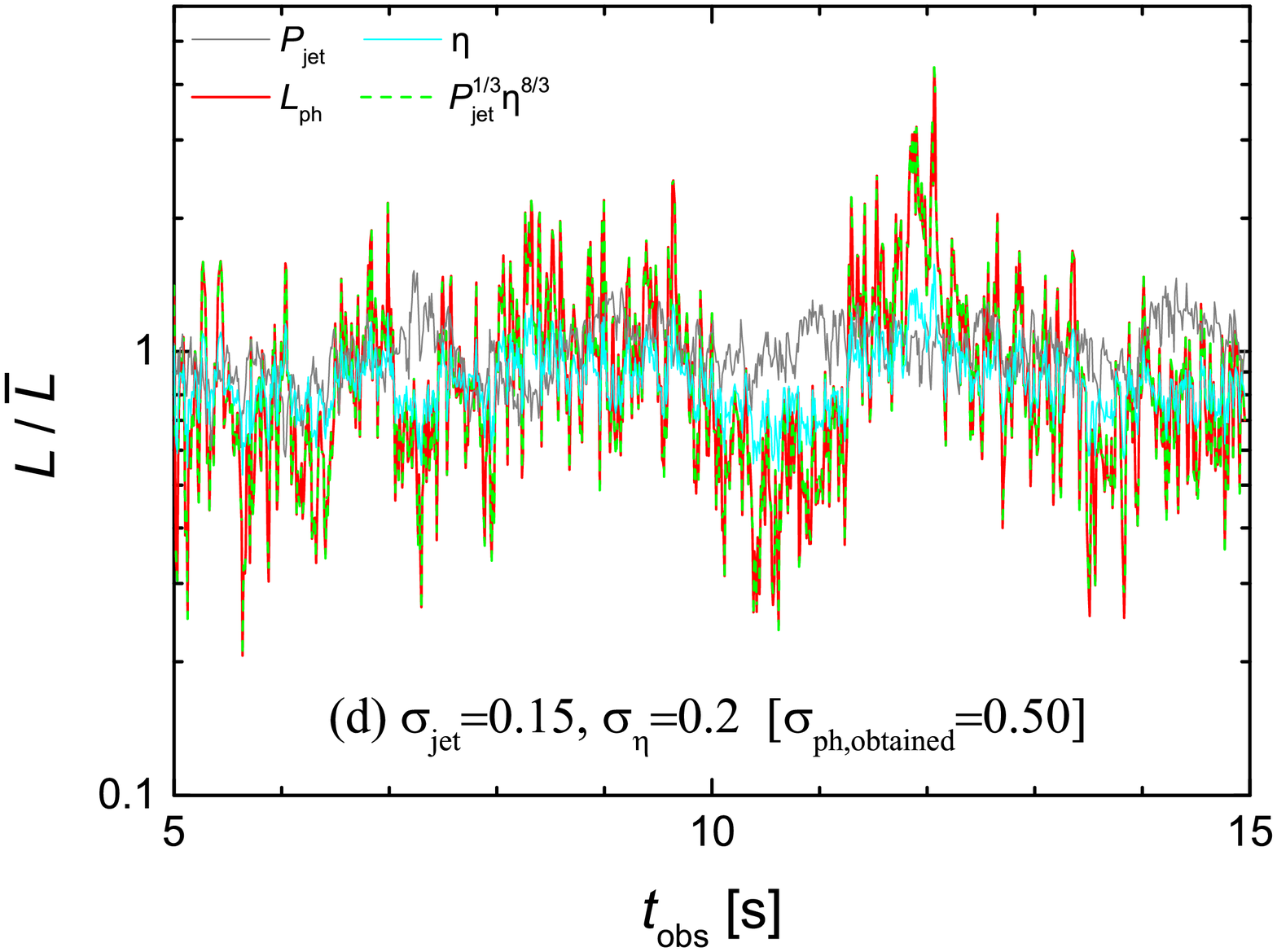} \\
\includegraphics[angle=0,width=0.5\textwidth]{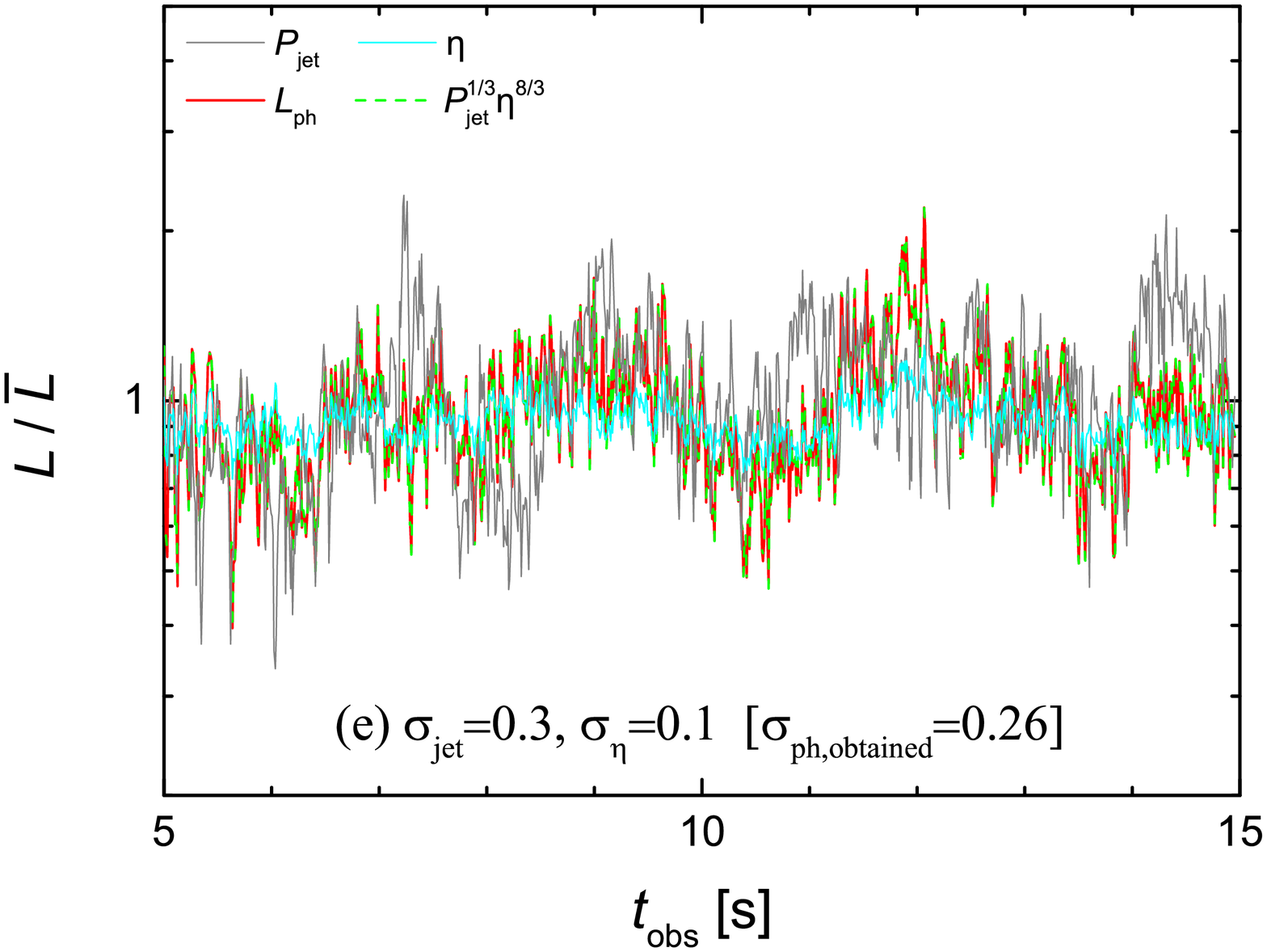} &
\includegraphics[angle=0,width=0.5\textwidth]{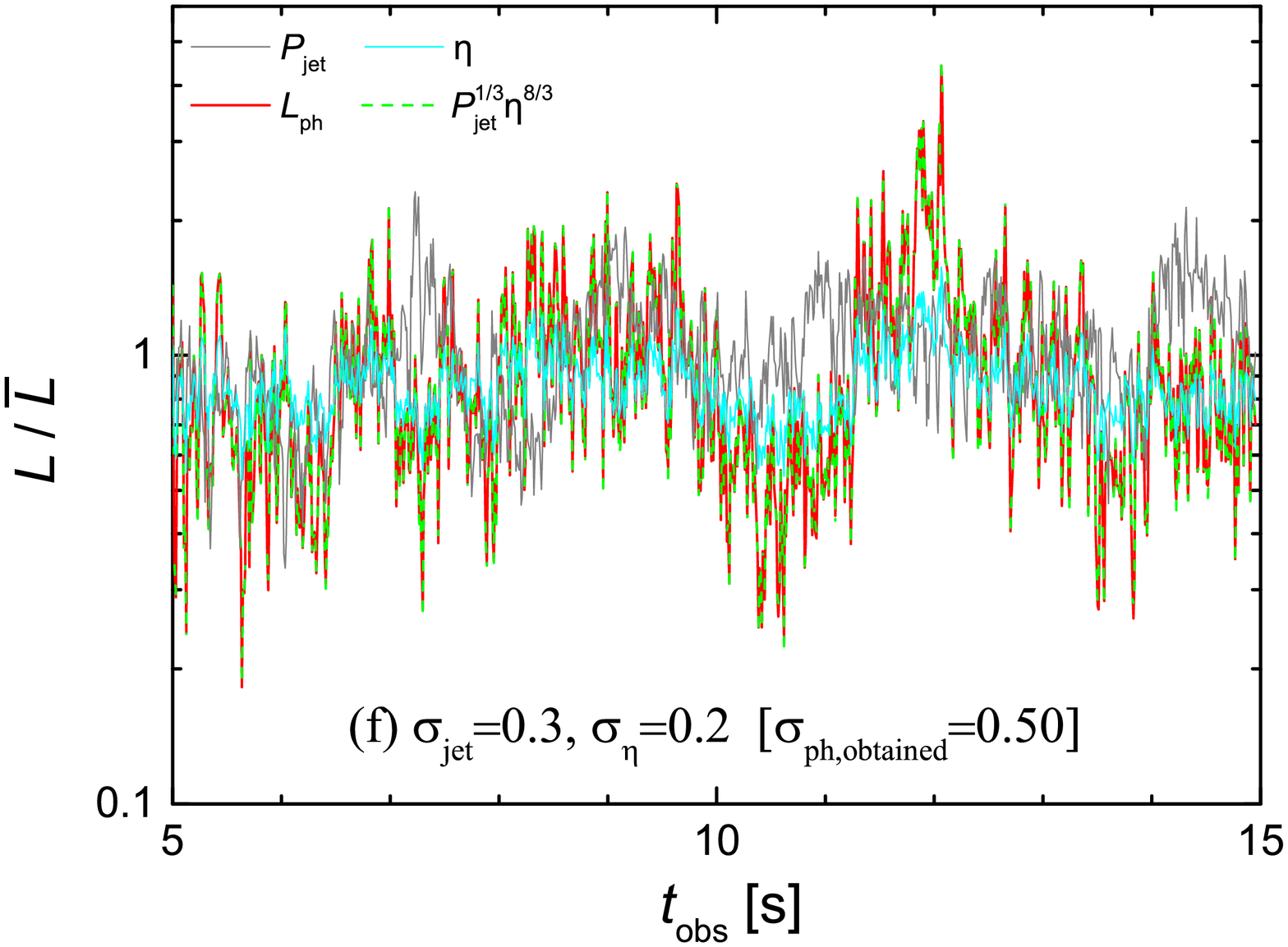} \\
\end{tabular}
\caption{Light curves of $P_{\rm jet}$ (gray lines), $\eta$ (cyan lines),
$L_{\rm ph}$ (red lines), and $P_{\rm jet}^{1/3}\eta^{8/3}$ (green dashed lines),
where jets with different $(\sigma_{\rm jet}, \sigma_{\eta})$ are studied.
The values of $\sigma_{\rm jet}$, $\sigma_{\eta}$,
and $\sigma_{\rm ph, obtained}$ (estimated based on $L_{\rm ph}$) are shown in the bottom of each panel.
The blue lines in the upper panels plot the variability of $F_{E}$ with $E=10^{-3}E_{\rm p}$.
}
\label{fig:L2}
\end{figure}

\clearpage
\begin{figure}
\centering
\begin{tabular}{cc}
\includegraphics[angle=0,width=0.5\textwidth]{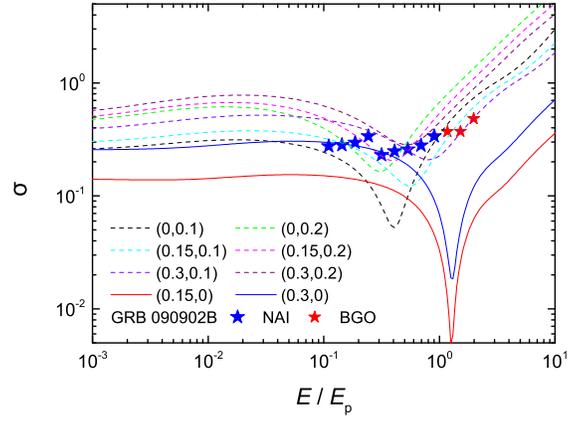}
\end{tabular}
\caption{Dependence of $\sigma_{_E}$ on $E$ with $E_{\rm{p}}$ being the peak photon energy of the $\nu F_\nu$ radiation spectrum.
Here, the symbol ``$\bigstar$'' represents the data from GRB~090902B.
}
\label{fig:sample1}
\end{figure}

\clearpage
\begin{figure}
\centering
\begin{tabular}{cc}
\includegraphics[angle=0,width=0.5\textwidth]{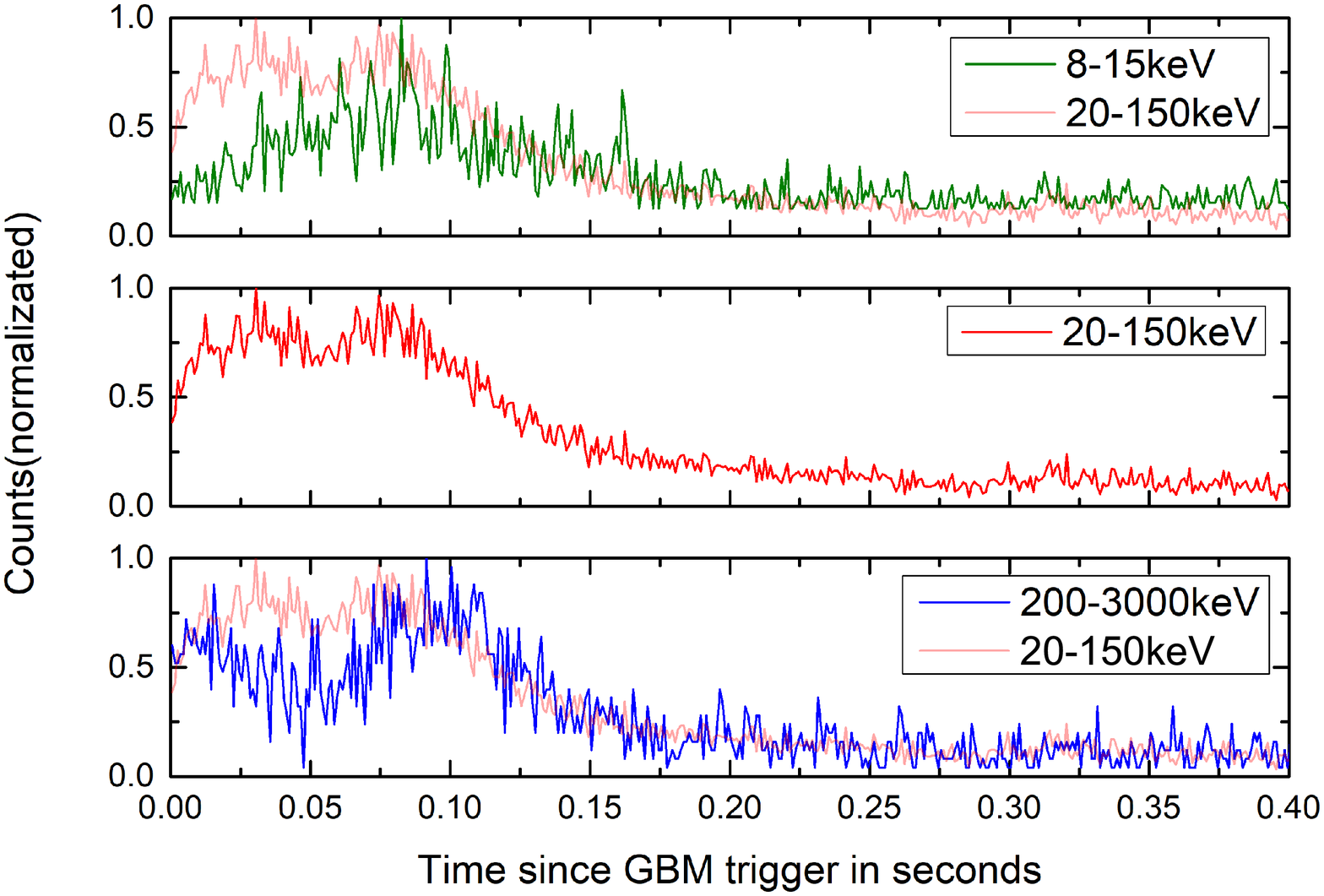} &
\includegraphics[angle=0,width=0.5\textwidth]{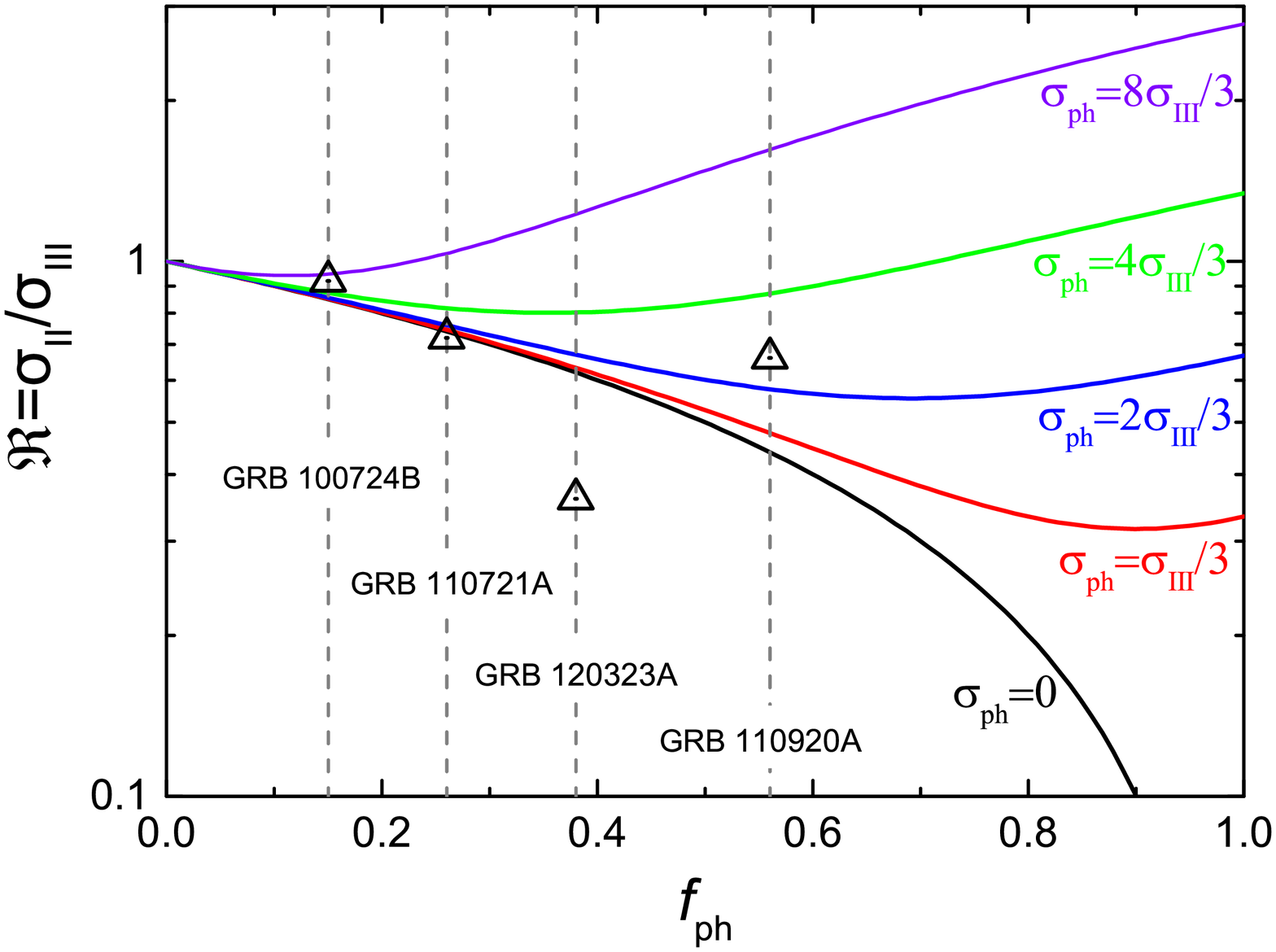}\\
\end{tabular}
\caption{\emph{Left Panel}: light curves of GRB~120323A in bands I (green line), II (red lines),
and III (blue line), where the Light curve is normalized with its maximal value;
\emph{Right Panel}: The dependence of $\Re=\sigma_{_{\rm II}}/\sigma_{_{\rm III}}$ on $f_{\rm ph}$
with different value of $\sigma_{\rm ph}/\sigma_{\rm III}$.
The symbol ``$\Delta$'' represents the data of our bursts.
}
\label{fig:sample2}
\end{figure}

\begin{table}[]
\centering
\caption{Values of $\sigma$, $\Re$, and $f_{\rm th}$ for our selected bursts.}
\begin{tabular}{cccccc}
\hline \hline
Burst & $\sigma_{_{\rm I}}$ & $\sigma_{_{\rm II}}$ & $\sigma_{_{\rm III}}$ & $\Re$ & $f_{\rm ph}$\\
\hline
GRB~100724B & 0.07 & 0.08  & 0.09 & 0.92 & 0.15     \\
GRB~110721A & 0.25 & 0.16  & 0.22 & 0.72 & 0.26     \\
GRB~110920A & 0.07 & 0.05  & 0.08 & 0.66 & 0.56     \\
GRB~120323A & 0.28 & 0.10  & 0.28 & 0.36 & 0.38     \\
 \hline
\end{tabular}
\label{table:1}
\end{table}
\clearpage
\bibliography{E:/Ref/bibliography}


\end{document}